# How fast can a quantum computer search?

*Lov K. Grover, lkgrover@bell-labs.com*[1]

This paper gives a simple proof of why a quantum computer, despite being in all possible states simultaneously, needs at least $0.707\sqrt{N}$ queries to retrieve a desired item from an unsorted list of $N$ items. The proof is refined to show that a quantum computer would need at least $0.785\sqrt{N}$ queries. The quantum search algorithm needs precisely this many queries.

**Introduction** Consider the problem of finding an item in an unsorted list of $N$ items that satisfies a desired property. No classical algorithm can guarantee to find this item in fewer than $N$ steps. Quantum mechanical systems can simultaneously be in multiple states and there is no clear bound on how fast a quantum computer could search this list. It was proved in 1994, through subtle properties of unitary transforms, that any quantum mechanical system would need at least $O(\sqrt{N})$ steps for such a search. Two years later in 1996, without being aware of the result proved earlier, I developed a quantum mechanical algorithm that indeed searched the list in $O(\sqrt{N})$ steps. Somewhat surprisingly, this algorithm was considerably simpler than the 1994 proof of the $O(\sqrt{N})$ step bound. In the last 3 years there have been several attempts to elaborate and improve the bound, this has led to the development of different concepts and approaches.

[bbbv] the original 1994 paper mentioned above, this proved the first $O(\sqrt{N})$ step bound.

[bbht] this derived an $O(\sqrt{N})$ step bound in the context of search problems, showed that the search algorithm [gro] was within 10% of optimal.

[zal] proved that the search algorithm [gro] was *precisely* optimal.

[pol] developed a new technique to lower bound quantum mechanical complexity of various problems through expansions of the state vectors in polynomials. Used this to prove the $O(\sqrt{N})$ step bound for the search problem.

[mos], [pre] and [ozh] give their own proofs of the $O(\sqrt{N})$ step bound. The proofs in all these papers are very involved. This paper is an attempt to develop a simple proof of the bound that will highlight the intrinsic reason for the bound.

**The three results** In order to see why this limit comes about, consider an ensemble of $N$ quantum computers each of which needs to return a different item in the list as the solution to the search problem. Each of the $N$ quantum computers takes the same initial state vector and through a different dynamic evolution converts it into a different final state vector - finally through a measurement, the desired item in the list is determined.

Different state vectors in $N$-dimensional space can be resolved only if they are mutually orthogonal - in order to be orthogonal they have to be *spread out*. Theorem 1 quantitatively shows that they should be spread out by $2N$ units. Theorem 2 shows that the spread after $t$ queries can be at most $4t^2$. This immediately shows that in order to distinguish between the various possibilities

1. This material is based upon work supported in part by the U. S. Army Research Office under contract no. DAAG55-98-C-0040.

will take at least $0.707\sqrt{N}$ queries. Theorem 3 refines the proof to show that the spread after $t$ queries can be only $4N\sin^2\frac{t}{\sqrt{N}}$ and the number of queries required will therefore need to be at least $0.785\sqrt{N}$ (the search algorithm requires precisely $0.785\sqrt{N}$ queries).

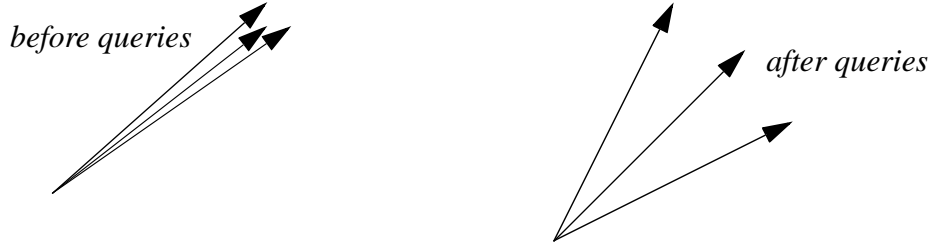

**Figure - Initially the ensemble of state vectors corresponding to the *N* different possibilities are parallel. As a result of the queries, they *spread* out. In order to be resolved they have to be spread out by *2N* units (theorem 1).**

**Theorem 1** - Consider any $N$ orthonormal vectors denoted by $\phi_1, \phi_2, \ldots, \phi_N$. Let $\phi_0$ be *any* reference unit vector. Then $\sum_{\alpha=1}^{N} |\phi_\alpha - \phi_0|^2 \geq (2N - 2\sqrt{N})$.

**Proof** - Since the $N$ unit vectors $\phi_1, \phi_2, \ldots, \phi_N$ are orthogonal, we can choose a frame with coordinate axes along these vectors. Following standard Dirac notation, denote the components of the unit vector $\phi_0$ along these axes by $\langle\phi_1|\phi_0\rangle, \langle\phi_2|\phi_0\rangle, \ldots, \langle\phi_N|\phi_0\rangle$. Therefore $\sum_{\alpha=1}^{N} |\phi_\alpha - \phi_0|^2$ becomes $\sum_{\alpha=1}^{N} \langle\phi_\alpha - \phi_0|\phi_\alpha - \phi_0\rangle$ which is equal to $\sum_{\alpha=1}^{N} (\langle\phi_\alpha|\phi_\alpha\rangle + \langle\phi_0|\phi_0\rangle - \langle\phi_\alpha|\phi_0\rangle - \langle\phi_0|\phi_\alpha\rangle)$, since $\phi_0$ & $\phi_\alpha$ are normalized, this becomes $\left(2N - \sum_{\alpha=1}^{N} (\langle\phi_\alpha|\phi_0\rangle + \langle\phi_\alpha|\phi_0\rangle^*)\right)$. The maximum value of $\sum_{\alpha=1}^{N} (\langle\phi_\alpha|\phi_0\rangle + \langle\phi_\alpha|\phi_0\rangle^*)$, subject to the constraint that $\phi_0$ is normalized, is attained when all $\langle\phi_\alpha|\phi_0\rangle$ are real and equal to $\frac{1}{\sqrt{N}}$. The corresponding value of $\sum_{\alpha=1}^{N} |\phi_\alpha - \phi_0|^2$ becomes $(2N - 2\sqrt{N})$.

The operation that changes the relative spread of the $N$ vectors is a selective phase rotation of the desired item (implemented through a query by the technique described in [bbht] & [stoc]). Other intermediate unitary transformations are rigid rotation of the ensemble in the state space and do not change the relative spread between the state vectors corresponding to the various possible solutions (these are required to prepare the system so that subsequent queries can induce greater increase in the spread). Denote the state vectors corresponding to the $N$ possibilities, just before the $t^{th}$ query, by $\psi_1(t-1), \psi_2(t-1), \ldots \psi_N(t-1)$. Theorem 2 shows by considering the mechanics of the query that in $t$ queries the spread increases to at most $4t^2$.

In order to quantify the spread, the reference state vector (which is denoted by $\phi_0$ in theorem 1) will be chosen to be the one which undergoes the same unitary transformations as the ensemble, but *without* any queries - just before the $t^{th}$ query, this reference vector is denoted by $\psi_0(t-1)$. The deviation of the $\alpha^{th}$ state vector from the reference vector, just before the $t^{th}$ query is denoted by $\Delta_\alpha(t-1)$ and defined as $\Delta_\alpha(t-1) \equiv |\psi_\alpha(t-1) - \psi_0(t-1)|^2$. The total spread of the system just before the $t^{th}$ query is defined by $\Delta(t-1)$:

$$\Delta(t-1) \equiv \sum_{\alpha=1}^{N} \Delta_\alpha(t-1) \equiv \sum_{\alpha=1}^{N} |\psi_\alpha(t-1) - \psi_0(t-1)|^2.$$

According to theorem 1, the $N$ state vectors $\psi_1(t), \psi_2(t), \ldots \psi_N(t)$ clearly cannot be resolved if the spread $\Delta(t)$ be less than $2N - 2\sqrt{N}$.

**Theorem 2** - The spread $\Delta(t)$, after $t$ queries, is upper bounded by: $\Delta(t) \leq 4t^2$.

**Proof** - In order to bound the increase of $\Delta_\alpha(t-1)$ due to $t^{th}$ query, consider the effect of the $t^{th}$ query on the state vector $\psi_\alpha(t-1)$. The query can alter only the phase of the $\alpha^{th}$ component of this vector.

The difference between $\Delta_\alpha(t)$ & $\Delta_\alpha(t-1)$ is first upper bounded. As in theorem 1, following standard Dirac notation, denote the $\alpha^{th}$ components of $\psi_\alpha(t-1)$ & $\psi_0(t-1)$ by $\langle\alpha|\psi_\alpha(t-1)\rangle$ & $\langle\alpha|\psi_0(t-1)\rangle$ respectively. The maximum change in $\Delta_\alpha(t-1)$ that the query can produce, happens when the $\alpha^{th}$ components of $\psi_\alpha(t-1)$ & $\psi_0(t-1)$ are initially parallel, but as a result of the query operation become anti-parallel.

$$\Delta_\alpha(t) - \Delta_\alpha(t-1) \leq (|\langle\alpha|\psi_\alpha(t-1)\rangle| + |\langle\alpha|\psi_0(t-1)\rangle|)^2 - (|\langle\alpha|\psi_\alpha(t-1)\rangle| - |\langle\alpha|\psi_0(t-1)\rangle|)^2$$
$$= 4|\langle\alpha|\psi_\alpha(t-1)\rangle||\langle\alpha|\psi_0(t-1)\rangle|$$
$$= 4|(\langle\alpha|\psi_\alpha(t-1)\rangle - \langle\alpha|\psi_0(t-1)\rangle) + \langle\alpha|\psi_0(t-1)\rangle||\langle\alpha|\psi_0(t-1)\rangle|$$

Using the inequality $|a+b| \leq |a| + |b|$ to $|(\langle\alpha|\psi_\alpha(t-1)\rangle - \langle\alpha|\psi_0(t-1)\rangle) + \langle\alpha|\psi_0(t-1)\rangle|$:

$$\leq 4|\langle\alpha|\psi_\alpha(t-1)\rangle - \langle\alpha|\psi_0(t-1)\rangle||\langle\alpha|\psi_0(t-1)\rangle| + 4|\langle\alpha|\psi_0(t-1)\rangle|^2$$

Using the inequality $4|a||b| \leq \frac{2}{\lambda}|a|^2 + 2\lambda|b|^2$ which holds for any positive $\lambda$:

$$\leq \frac{2}{\lambda}|\langle\alpha|\psi_\alpha(t-1)\rangle - \langle\alpha|\psi_0(t-1)\rangle|^2 + 2\lambda|\langle\alpha|\psi_0(t-1)\rangle|^2 + 4|\langle\alpha|\psi_0(t-1)\rangle|^2$$

Summing over all $\alpha$, and using the following three relations:

$$|\langle\alpha|\psi_\alpha(t-1)\rangle - \langle\alpha|\psi_0(t-1)\rangle|^2 \leq \Delta_\alpha(t-1) \ \& \ \Delta(t) \equiv \sum_{\alpha=1}^{N}\Delta_\alpha(t) \ \& \ \sum_{\alpha=1}^{N}|\langle\alpha|\psi_0(t-1)\rangle|^2 = 1,$$

it follows: $\Delta(t) - \Delta(t-1) \leq \frac{2}{\lambda}\Delta(t-1) + (2\lambda + 4)$

Choosing $\lambda$ as $\sqrt{\Delta(t-1)}$, it follows that: $\Delta(t) - \Delta(t-1) \leq 4\sqrt{\Delta(t-1)} + 4$. By shuffling terms, and taking the square-root of both sides, this may be written as: $\sqrt{\Delta(t)} \leq (\sqrt{\Delta(t-1)} + 2)$ which implies that $\sqrt{\Delta(t)} \leq 2t$ or equivalently $\Delta(t) \leq 4t^2$.

By slightly refining the calculations in theorem 2, it is possible to show that the quantum search algorithm is asymptotically optimal (as in [zal]). Indeed it is possible to improve the bound of theorem 2 to prove: $\Delta(t) \leq 4N\sin^2\frac{t}{\sqrt{N}}$. For $t \ll \sqrt{N}$, $\sin\frac{t}{\sqrt{N}} \approx \frac{t}{\sqrt{N}}$ and the upper bound for $\Delta(t)$ becomes $4t^2$ which is the same as in theorem 2. However, when $t$ is $O(\sqrt{N})$, $\sin^2\frac{t}{\sqrt{N}} < \frac{t^2}{N}$, and the bound for $\Delta(t)$ is tighter. In order to make $\Delta(t)$ equal to $2N$, $t$ will have to be at least $\frac{\pi\sqrt{N}}{4}$. Therefore the number of oracle accesses required will be at least $0.785\sqrt{N}$. The quantum search algorithm requires precisely this many oracle queries.

The way to refine the calculation of theorem 2 is by improving the bound: $|\langle\alpha|\psi_\alpha(t-1)\rangle - \langle\alpha|\psi_0(t-1)\rangle|^2 \leq \Delta_\alpha(t-1)$ which was used in the last step of the calculation. The additional idea we use is that since $\psi_\alpha(t-1)$ & $\psi_0(t-1)$ are both unit vectors, and one of their components differs by a certain amount, the other components will also have to differ. Therefore it is possible to obtain a slightly tighter lower bound for $\Delta_\alpha(t-1)$. This is elaborated in the following lemma -

**Lemma -** Consider two unit vectors $\phi_1$ and $\phi_2$ in $N$-dimensional complex space. In case $\sum_{\alpha=1}^{N}|\langle\alpha|\phi_1\rangle - \langle\alpha|\phi_2\rangle|^2 = 2 - 2\cos\theta$, then it follows that the magnitude of any of the $N$ components which are known to be of the same sign, say the $\beta^{th}$ component, can differ by at most $\sin^2\theta$, i.e. $(|\langle\beta|\phi_1\rangle| - |\langle\beta|\phi_2\rangle|)^2 \leq \sin^2\theta$.

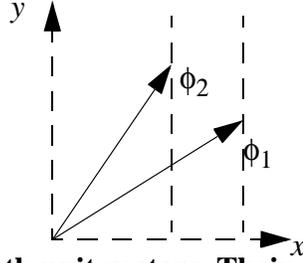

**Figure - $\phi_1$ & $\phi_2$ are both unit vectors. Their *y* component differs by a specified amount, then their *x* component will also have to differ by a related amount (lemma).**

**Idea of proof -** The main idea behind this lemma is that since the two vectors, $\phi_1$ and $\phi_2$, are both unit vectors, they have the same length. Therefore if one of the components differ, some of the remaining components will have to differ too.

As shown in the following figure, in case the *y* components differ by $\Delta y$, the *x* components will also have to differ by a related amount $\Delta x$. The difference $\Delta$ between the two vectors will be $\Delta^2 \geq (\Delta x)^2 + (\Delta y)^2$. In theorem 2, we used just the weaker inequality $\Delta^2 > (\Delta y)^2$.

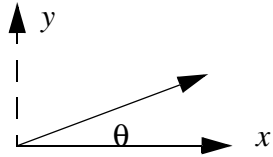

If $\Delta y^2 = \sin^2\theta$, then the minimum separation between the two unit vectors is when the first unit vector has zero *y* component and the other has a *y* component equal to $\sin \theta$.

The separation then becomes $\Delta y^2 + \Delta x^2$.

which is equal to $\sin^2\theta + (1-\cos\theta)^2 = (2 - 2\cos\theta)$.

**Figure - In case one component of two unit vectors (say *y*) is known to differ, then some other component (*x*) will also have to differ.**

**Theorem 3** - The spread $\Delta(t)$ is upper bounded by $\Delta(t) \leq 4N\sin^2\left(\frac{t}{\sqrt{N}} + \frac{t}{2N}\right)$.

**Proof** - The first portion of the theorem is the same as theorem 2. As in theorem 2, we observe that the maximum increase in $\Delta_\alpha(t-1)$ that the query can produce, happens when the $\alpha^{th}$ components of $\psi_\alpha(t-1)$ & $\psi_0(t-1)$ are initially parallel, but as a result of the query operation become anti-parallel. This leads to the following bound which holds for any positive $\lambda$.

$$\Delta_\alpha(t) - \Delta_\alpha(t-1) \leq \frac{2}{\lambda}\bigl|\,|\langle\alpha|\psi_\alpha(t-1)\rangle| - |\langle\alpha|\psi_0(t-1)\rangle|\,\bigr|^2 + 2\lambda|\langle\alpha|\psi_0(t-1)\rangle|^2 + 4|\langle\alpha|\psi_0(t-1)\rangle|$$

The remainder of the proof proves a bound tighter than theorem 2 by using the lemma proved just before theorem 3. Using this lemma, we upper bound $\bigl|\,|\langle\alpha|\psi_\alpha(t-1)\rangle| - |\langle\alpha|\psi_0(t-1)\rangle|\,\bigr|^2$ in terms of the difference of the two vectors $\psi_\alpha(t-1)$ & $\psi_0(t-1)$. If $\Delta_\alpha(t-1) \equiv |\psi_\alpha(t-1) - \psi_0(t-1)|^2 = (2 - 2\cos\theta_\alpha)$, then

$\||\langle\alpha|\psi_\alpha(t-1)\rangle| - |\langle\alpha|\psi_0(t-1)\rangle|\|^2 \leq \sin^2\theta_\alpha$. Therefore:

if $\Delta(t-1) = \sum_{\alpha=1}^{N}(2-2\cos\theta_\alpha)$, then $\Delta(t) - \Delta(t-1) \leq \frac{2}{\lambda}\sum_{\alpha=1}^{N}\sin^2\theta_\alpha + 2\lambda + 4$.

It is easily seen that if $\sum_{\alpha=1}^{N}(2-2\cos\theta_\alpha)$ is fixed, then $\sum_{\alpha=1}^{N}\sin^2\theta_\alpha$ is maximized when all $\theta_\alpha$ are identical. Therefore if $\sum_{\alpha=1}^{N}(2-2\cos\theta_\alpha) \equiv (2N-2N\cos\Theta)$, then $\sum_{\alpha=1}^{N}\sin^2\theta_\alpha \leq N\sin^2\Theta$. Therefore: if $\Delta(t-1) = (2N-2N\cos\Theta_t)$, then $\Delta(t) - \Delta(t-1) \leq \frac{2}{\lambda}N\sin^2\Theta_t + 2\lambda + 4$.

Writing $(2N-2N\cos\Theta_t)$ as $4N\sin^2\frac{\Theta_t}{2}$ and choosing $\lambda$ as $\sqrt{N}\sin\Theta_t$ (which maximizes $\left(\frac{2}{\lambda}N\sin^2\Theta_t + 2\lambda + 4\right)$): if $\Delta(t-1) = 4N\sin^2\frac{\Theta_t}{2}$, then $\Delta(t) \leq \Delta(t-1) + 4\sqrt{N}\sin\Theta_t + 4$.

Now using the relations $\Delta(t-1) = 4N\sin^2\frac{\Theta_t}{2}$ and $\Delta(t) \leq \Delta(t-1) + 4\sqrt{N}\sin\Theta_t + 4$, we prove by induction that $\Delta(t) \leq 4N\sin^2\left(\frac{t}{\sqrt{N}} + \frac{t}{2N}\right)$.

First observe that $\Delta((t-1)=0) = 0$ and from the relation $\Delta(t-1) = 4N\sin^2\frac{\Theta_t}{2}$ it follows that when $(t-1) = 0$, then $\Theta_t = 0$.

Substituting $\Delta(t-1) = 4N\sin^2\frac{\Theta_t}{2}$ in the inequality $\Delta(t) \leq \Delta(t-1) + 4\sqrt{N}\sin\Theta_t + 4$ it follows that $\Delta(t) \leq 4N\sin^2\frac{\Theta_t}{2} + 4\sqrt{N}\sin\Theta_t + 4$. Using elementary calculus, it can be shown:

$4N\sin^2\frac{\Theta_t}{2} + 4\sqrt{N}\sin\Theta_t + 4 \leq 4N\sin^2\left(\frac{\Theta_t}{2} + \frac{1}{\sqrt{N}} + \frac{1}{2N}\right)$, for all values of $\Theta_t$ in the range $\left[0, \frac{\pi}{2}\right]$.

Therefore: $\Delta(t) \leq 4N\sin^2\left(\frac{\Theta_t}{2} + \frac{1}{\sqrt{N}} + \frac{1}{2N}\right)$. This shows that the parameter $\frac{\Theta_t}{2}$ increases by at most $\left(\frac{1}{\sqrt{N}} + \frac{1}{2N}\right)$ at each time step. It follows by induction that $\Delta(t) \leq 4N\sin^2\left(\frac{t}{\sqrt{N}} + \frac{t}{2N}\right)$.

Note that the exact variation of $\Delta(t)$ for quantum search is given by $\Delta(t) = 4N\sin^2\left(t\sin^{-1}\left(\frac{1}{\sqrt{N}}\right)\right)$ and the upper bound proved in this section comes very close to this.

**Acknowledgements** - Charlie Bennett, Norm Margolus, Christof Zalka and particularly Apoorva Patel suggested various improvements.